\documentclass[a4paper,12pt]{article}
\pdfoutput=1
\usepackage{slashed}
\usepackage[affil-sl,auth-sc]{authblk}
\usepackage{amssymb,amsmath,amsbsy}
\usepackage{bbm}
\usepackage{mathrsfs}
\usepackage{mathbbol}
\usepackage{bm}
\usepackage{appendix}
\usepackage{hyperref}
\usepackage[top=1.2in, bottom=1.2in,left=0.9in,includefoot]{geometry}
               
\usepackage{graphicx}
\usepackage{cite}
\usepackage{float}
\usepackage{caption}
\usepackage{wasysym}
\usepackage{amsthm}
\usepackage[utf8]{inputenc}
\usepackage[english]{babel}
\newif\ifContLineTwo
\newif\ifContLineThree
\def\conC#1{\vbox{\ialign{##\crcr
  \ifContLineThree\hrulefill\else\vphantom{\hrulefill}\fi\crcr
  \noalign{\kern3.2pt\nointerlineskip}
  \ifContLineTwo\hrulefill\else\vphantom{\hrulefill}\fi\crcr
  \noalign{\kern3.2pt\nointerlineskip}
  \ifContLineOne\hrulefill\else\vphantom{\hrulefill}\fi\crcr
  \noalign{\nointerlineskip}
  $\hfil\textstyle{\vbox to 14pt{}#1}\hfil$\crcr}}}
\def\DrawLeg#1#2{
  \kern-.2pt              
  \dimen2 =#1             
  \advance\dimen2 by 2pt  
  \dimen3 = 10.6pt        
  \dimen4 =3.6pt          
  \advance\dimen3 by -\dimen2 
  \multiply\dimen4 by #2
  \advance\dimen3 by \dimen4
  \raise\dimen2 \hbox{\vrule height\dimen3 width .4pt} 
  \kern-.2pt}             
\def\begC#1#2{\setbox0 =\hbox{$\textstyle{#2}$}
  \dimen0=.5\wd0 \dimen1=\ht0
  \conC{\hskip\dimen0}
  \count255=#1
  \ifnum\count255 =1 \ContLineOnetrue\else
  \ifnum\count255 =2 \ContLineTwotrue\else
  \ifnum\count255 =3 \ContLineThreetrue\fi\fi\fi
  \DrawLeg{\dimen1}{\count255}
  \conC{\hskip\dimen0}
  \kern-\dimen0\kern-\dimen0 \box0}
\def\endC#1#2{\setbox0 =\hbox{$\textstyle{#2}$}
  \dimen0=.5\wd0 \dimen1=\ht0
  \conC{\hskip\dimen0}
  \count255=#1
  \ifnum\count255 =1 \ContLineOnefalse\else
  \ifnum\count255 =2 \ContLineTwofalse\else
  \ifnum\count255 =3 \ContLineThreefalse\fi\fi\fi
  \DrawLeg{\dimen1}{\count255}
  \conC{\hskip\dimen0}
  \kern-\dimen0\kern-\dimen0 \box0}

\newcommand{\pythia}{{\tt PYTHIA8-8.2.19}}
\newcommand{\madgraph}{\tt MadGraph5\_aMC@NLO-2.4.3}

\newcommand{\hepmc}{\tt HepMC-2.06.09}
\newcommand{\delphes}{\tt Delphes-3.3.3}

\def\lsim{\buildrel{\scriptscriptstyle <}\over{\scriptscriptstyle\sim}}
\def\gsim{\buildrel{\scriptscriptstyle >}\over{\scriptscriptstyle\sim}}

\newcommand{\BRAG}{$ \text{Br} \left( A_1 \rightarrow \gamma \gamma \right)$}

\newcommand{\bea}{\begin{eqnarray}}
\newcommand{\eea}{\end{eqnarray}}
\newcommand{\nn}{\nonumber\\}

\usepackage{color}

\definecolor{orange}{rgb}{0.9,0.2,0}
\definecolor{brown}{rgb}{0.7,0.3,0.2}
\definecolor{fuxia}{rgb}{1,0,1}
\definecolor{skyblue}{rgb}{0,0.1,0.9}
\definecolor{violetred}{rgb}{0.8,0.13,0.56}
\definecolor{deeppink}{rgb}{1.00,0.08,0.5}
\definecolor{pink}{rgb}{1.00,0.75,0.80}
\definecolor{orchid}{rgb}{0.85,0.44,0.84}
\definecolor{lightpink}{rgb}{1.00,0.71,0.76}
\definecolor{bluish}{rgb}{0,0.6,0.8}
\numberwithin{equation}{section}
\title{\bf Detection prospects of light pseudoscalar Higgs boson at the LHC}
\author{\small Monoranjan Guchait \thanks{guchait@tifr.res.in}~}
\author{~\small Aravind H. Vijay \thanks{arvind.vijay@tifr.res.in}}
\author{~\small Jacky Kumar\thanks{jka@tifr.res.in}}
\affil{\small Department of High Energy Physics, \\
Tata Institute of Fundamental Research, \\
 Homi Bhabha Road, Mumbai-400005, India}

\def \PMET{E{\!\!\!/}_T}
\def \MET{E{\!\!\!/}_T}
\def\invfb{\rm fb^{-1}}
\def\hsm{\rm H}

\def\t1 {\widetilde {t_1}}
\def\N1{\widetilde \chi_1^0}
\def\N2{\widetilde \chi_2^0}
\def\N3{\widetilde \chi_3^0}
\def\N0{\widetilde \chi^0}
\def\C1{\widetilde \chi_1^{\pm}}
\def\mst1 {m_{\t1}}
\def\br {\begin{eqnarray}}
\def\er {\end{eqnarray}}

\def\lsim{\buildrel{\scriptscriptstyle <}\over{\scriptscriptstyle\sim}}
\def\gsim{\buildrel{\scriptscriptstyle >}\over{\scriptscriptstyle\sim}}

\date{}

\begin{document}
\maketitle
\begin{abstract}
The discovery potential of light pseudo scalar Higgs boson  
for the mass range 10-60 GeV is explored.
In the context of the next-to-minimal supersymmetric 
standard(NMSSM) model, the branching fraction of light 
pseudo scalar Higgs boson
decaying to a pair of photon can be quite large.     
A pair of light pseudo scalar Higgs boson produced indirectly 
through the standard model Higgs boson decay yields 
multiple photons in the final state and the corresponding 
production rate is restricted by ATLAS data.
Discussing the impact of this constraint in the NMSSM, 
the detection prospects of light pseudoscalar  
Higgs boson in 
the channel consisting of at least three photons, a lepton 
and missing transverse energy are reported.
It is observed that the possibilities of finding the 
pseudoscalar Higgs boson for the above mass range
are promising for an integrated luminosity
${\cal L}$=100$~\rm \invfb$ with moderate significances,
which can reach to more than 5$\sigma$  
for higher luminosity options.
\end{abstract}
\vskip .5 true cm
\newpage
\section{Introduction}

The discovery of the Higgs boson of mass 125~GeV 
\cite{Aad:2012tfa,Chatrchyan:2012xdj} 
at the Large Hadron collider(LHC) by both the CMS and ATLAS 
experiments opens up a new window to study the 
physics beyond standard model (BSM). The current precision 
measurements of various properties of the 
Higgs boson, in particular couplings with fermions and gauge bosons 
indicate that it is indeed the candidate for the 
Standard model (SM) Higgs~\cite{Khachatryan:2016vau}.
However, the possibility of interpreting it as the candidate of
a BSM is not also ruled out completely.
For example, the minimal supersymmetric standard
model(MSSM), a candidate for BSM theories with two   
Higgs doublets leading to five physical Higgs boson states - 
h, H, A and $\rm H^\pm$, where the first two are the CP-even,
and others are the CP-odd and charged Higgs boson respectively, offers 
its lightest CP-even Higgs boson($\rm h$) as the candidate for the SM-like 
Higgs boson.
Theoretically, in the MSSM, the mass of the Higgs boson, 
precisely the lightest state h, is connected with other
sparticles strongly, in particular, with the third generation of squarks,
through higher order corrections which enhance its tree 
level mass substantially. 
Consequently, the corresponding region of the 
parameter space is constrained owing to 
the larger mass of the Higgs boson. 
For instance, Higgs mass of 125 GeV requires, either a heavier mass of 
top squarks or a very large mixing in the top squark sector predicting lighter
top squarks\cite{Hall:2011aa,Arbey:2011ab}. 
Furthermore, the non-minimal  
variations of the MSSM with an extended Higgs sector, 
such as the next to Minimal supersymmetric standard model(NMSSM)\cite{Fayet:1974pd,Ellis:1988er,doi:10.1142/S0217751X89001448,Ellwanger:2009dp} 
with an additional Higgs singlet field (S) in addition to two Higgs 
doublets, resulting in seven physical Higgs bosons,
also can offer one of its CP even Higgs boson as the candidate 
for the SM-like Higgs with 
mass 125 GeV.
Interestingly, in this case, one does not require so much fine tuning of 
the parameter space like the MSSM to accommodate the 125 GeV SM-like Higgs 
Boson~\cite{Hall:2011aa,Cao:2012fz}. 
In general, two Higgs doublet models of various types 
contain potential features to 
interpret one of its CP even Higgs boson as the SM-like in the alignment limit 
along the direction of vacuum expectation values(VEV) 
of doublets~\cite{Haber:2006ue,Bernon:2015qea,Bernon:2015wef}. 
Needless to say, that in all these cases, the BSM with extended Higgs 
sector predicts multiple Higgs boson states in addition to the 
SM-like one. Perhaps,  
discovery of an extra non-SM like Higgs boson might be one of the unambiguous 
signal to confirm the new physics model.

With this motivation, looking for the non-SM like Higgs boson at the LHC 
in the framework of the NMSSM has received a lot of attention 
since the discovery of the 125 GeV Higgs boson~\cite{Djouadi:2008uw,Heinemeyer:2011aa,King:2012is,Agashe:2012zq,King:2012tr,Vasquez:2012hn,Barbieri:2013hxa,Badziak:2013bda,Cao:2013gba,Christensen:2013dra,Guchait:2015owa,Domingo:2015eea,Guchait:2016pes,Kumar:2016vhm,Kozaczuk2015}.
The enlarged Higgs sector of the NMSSM interpreting  
one of the CP even state as the SM-like Higgs, predicts lighter
Higgs boson states, even much lower than 125 GeV for a wide
region of parameters~\cite{Guchait:2015owa}.
These lighter Higgs bosons are not ruled out by any past
experiments because of their suppressed couplings with gauge bosons 
and fermions. Many phenomenological studies exist in the 
literature regarding the searches of these lighter Higgs boson 
states at the LHC in the context of the NMSSM (see for details   
ref.~\cite{Ellwanger:2011sk} and references there in). 
On the experimental side, searching for these lighter
Higgs boson states at the LHC is also one of the
focused area since the discovery of the 125 GeV Higgs boson.
For instance, CMS and ATLAS experiments published results on the
searches of light Higgs bosons for various final states.
The non observation of any signal event predicts the model 
independent exclusion of the Higgs production rate  
corresponding to that final state, for a given mass of the 
Higgs boson~\cite{Chatrchyan:2012am,Veelken:1953441,Khachatryan:2015nba,Aad:2015bua}
Recasting these limits, possibly the parameter space of several models 
with extended Higgs can be constrained~\cite{Cacciapaglia:2016tlr,
Aggleton:2016tdd,Bhatia:2017ttp}.

In this present study, we explore the discovery  potential of light pseudo 
scalar Higgs boson($\rm A_1$) of mass less than $\rm m_{\hsm}/{2} $
at the LHC Run 2 experiment with the center of mass energy $\sqrt{S}=$13~TeV,
where H represents the SM Higgs.
As noted earlier, the direct production of these light 
singlet like Higgs boson state is suppressed due to the 
tiny couplings with the fermions.
Alternatively, we consider the production of light $\rm A_1$
indirectly via the decay of the SM Higgs boson as,
\br
\rm H \to A_1 A_1,
\label{eq:hsmdk}
\er
with the production of $\rm H$ via the standard mechanisms.
It is to be noted here that the total branching ratio of the SM Higgs to 
undetected decay modes $(\rm BR_{BSM})$
is restricted by Higgs data, and predicted 
to be~\cite{Khachatryan:2016vau},
\br
\rm BR_{\rm BSM}\lesssim 0.34 \ \ {\rm at} \ \ 95\% \ \ C.L.
\label{eq:bsmbr}
\er
Hence, the presence of non-SM decay modes of SM-like Higgs boson is not
completely ruled out.
Unlike the couplings of singlet like $\rm A_1$ with the fermions, 
the tree level Higgs-to-Higgs coupling, $\rm \hsm$-$\rm A_1$-$\rm A_1$ is 
not suppressed and widely varies with the model 
parameters~\cite{Ellwanger:2009dp}, to be discussed in the next section.  

The decay modes of light singlet like $\rm A_1$ are very interesting.
Due to the presence of finite fraction of doublet component in its
physical state, $\rm A_1$ dominantly decays via $\rm A_1 \to b\bar b$,
along with other sub-dominant channels such as, $\mu\mu, \tau\tau$.
However, surprisingly, it is observed that corresponding to a certain
kind of parameter space, the branching fraction (BR) of,
\br 
\rm A_1 \to \gamma\gamma, \nn
\label{eq:a12gg}
\er
can be as large as 100\%.
This large BR of $\rm A_1$ into this di-photon channel is a
novel feature of the NMSSM unlike the other 
SUSY models~\cite{Badziak:2013bda,Christensen:2013dra,Guchait:2015owa,Dermisek:2007yt,Ellwanger:2011aa}
Since experimentally, photon is a well reconstructed clean object, this
di-photon mode is expected to provide a robust signal of $\rm A_1$
at the LHC. Following this, we try to exploit this di-photon mode of
$\rm A_1$, to find its discovery potential at the LHC Run 2 experiment.
At the parton level, the production of a 
pair of $\rm A_1$ via the ~\eqref{eq:hsmdk}, and its subsequent decay 
following the ~\eqref{eq:a12gg}, results in four photons in the final 
state.  
Presence of multiple photons in the final state is of course, 
very encouraging, since the contamination due to the SM 
backgrounds is not expected to be severe~\cite{Moretti:2006sv,Arhrib:2006sx,Badziak:2013gla,Ellwanger:2015uaz}. In this present study, 
the production of the SM Higgs $\rm \hsm$ is considered in association with 
top pairs($t\bar t\hsm$), 
gauge bosons(W$\rm \hsm$, Z$\hsm$) 
and also the single top quark(t$\rm \hsm$).
In order to regulate the possible contamination 
due to the SM backgrounds, we demand at least one hard  
lepton arising from W or semileptonic decays of top quarks,
in the final state along with at least three photons.
The requirement of one lepton in the final state barring
us from using the $\hsm$ production via gluon-gluon
fusion, which is already analyzed and reported 
in the paper of Ref.~\cite{Chang:2006bw}. 

Hence, the production mechanism of the signal,
$\rm n_{\gamma} \gamma + \rm n_{\ell}\ell + \PMET$
($\ell = e,\mu$, ~$\rm n_\gamma \ge 3,~ \rm n_\ell \ge 1$)  
can be represented as, 
\br
{\rm pp} &\to& {\rm t \bar t \hsm},~ \rm W \hsm,~ \rm Z\hsm,~ \rm t\hsm  \nonumber \\
&\to& \rm n_{\gamma} \gamma + \rm n_{\ell}\ell + \PMET
\label{eq:signal}
\er
with $\rm \hsm$ and $\rm A_1$ decays given by \eqref{eq:hsmdk} and \eqref{eq:a12gg}
respectively.
The $\PMET$ arises due to the presence of neutrinos in the leptonic decays
of W and top quarks. 
The sources of various possible backgrounds are discussed 
in the later section.

Notice that the signal rate depends on the product of BRs as,
\br
\beta = \rm BR(\rm H \to A_1 A_1) \times BR(A_1 \to \gamma\gamma)^2. 
\label{eq:beta}  
\er
Interestingly, this product of BRs is found to be 
constrained by ATLAS data published recently~\cite{Aad:2015bua}. 
The ATLAS collaboration 
carried out searches for new phenomena in events with at least 
three photons at a center-of-mass energy 8~TeV with an integrated 
luminosity of 20.3$\invfb$. 
From the non observation of any excess events,
limits are set at 95\% C.L on the rate 
of events in terms of cross section multiplied by branching 
ratios~\cite{Aad:2015bua},
\br
\sigma \times \beta \lsim 10^{-3} \sigma_{SM}, 
\er
here $\sigma$ is the Higgs production cross section in new physics 
scenario, where as $\sigma_{\rm SM}$ is the same, but for the SM Higgs.  
The above constraint sets upper limit on $\beta$ as,
\br
\beta \lsim 10^{-3}, 
\label{eq:betalimit}
\er
provided the Higgs in the context of new physics phenomena is the SM like Higgs
boson of mass 125 GeV. 
Since our signal arises through the same decay channels,
\eqref{eq:hsmdk} and \eqref{eq:a12gg}, therefore the corresponding rate is
limited by the above constraint.
Moreover, the region of the parameter space, and hence the corresponding 
BRs of the decay channels, (\eqref{eq:hsmdk} and ~\eqref{eq:a12gg}), 
are also expected to be constrained, which are also investigated in this
study. Considering all these restrictions,
a detailed simulation of signal and backgrounds are carried out to find
the discovery potential of $\rm A_1$ at the LHC for few 
luminosity options.
  
We present our study as follows. Discussing the NMSSM model and decay modes
of $\rm A_1$ very briefly in Sec.~\eqref{sec:nmssm}, we present results of 
signal and background simulation 
in Sec.~\eqref{sec:sandb}.  Finally we summarize 
in Sec.~\eqref{sec:conclusion}.

\section{The NMSSM} 
\label{sec:nmssm}
It is instructive to discuss qualitatively some features of the NMSSM 
in the context of this present study.
As already mentioned, the Higgs sector in this model is extended by an 
additional Higgs field singlet under SM gauge transformation.
Historically, this model was motivated to address the $\mu$
problem~\cite{Kim:1983dt}, where $\mu$ is defined to be the Higgsino
mass parameter. Ideally, $\mu$ is free to take any value ranging
from EW scale to Planck scale which is not such a favorable scenario.
In order to cure this problem, the $\mu$ term is generated dynamically
through the vacuum expectation value (vev) of the singlet field.
In the $\mathbbm{Z}_3$ invariant NMSSM, the superpotential with 
two Higgs doublets($\rm {H_u, H_d}$) and a singlet(S) Higgs field 
is of the form:
\br
{ \rm \mathcal{W}_{NMSSM}} = { \rm \mathcal{ W}_{MSSM}|_{\mu =0}} + \lambda \rm  S  {H_u} {H_d} + \frac{1}{3}\kappa S^3, \label{eq:poteq}
\er
where $\lambda$ and $\kappa$ are the dimensionless couplings{\footnote 
{Perturbative nature of couplings 
requires, $\lambda^2 + \kappa^2 < 0.7$~\cite{Miller:2003ay}.}}
and $\mathcal{W}_{\rm MSSM}$ represents the superpotential in MSSM without 
$\mu$ term. Effect of Pecci-Quinn symmetry~\cite{Peccei:1977hh}
is avoided adding the extra $S^3$ term to the potential. Once
the singlet Higgs field receives vev ($v_s$), i.e, 
$\rm{\lambda S ~H_u~ H_d}$ $\sim$ $\rm \lambda v_s H_u ~H_d$, 
$\lambda \rm v_s$ appears to be effective $\mu$ parameter.
Alternate solutions also exist to address the $\mu$ 
problem~\cite{Giudice:1988yz,Nelson:2015cea}.
The presence of two Higgs doublets and one singlet field
lead to seven physical Higgs bosons states, such as,
3 CP-even Higgs 
states $\rm H_1, ~H_2, ~H_3$ (assuming $\rm m_{H_1} < m_{H_2} < m_{H_3}$),
2 CP-odd states $\rm A_1,~A_2 $ (assuming $\rm m_{A_1} < m_{A_2}$)
and two charged Higgs bosons~\cite{Ellwanger:2009dp}.
Contribution due to the additional singlet-doublet mixing 
term $\lambda \rm {S H_u H_d}$, enhances the tree level
Higgs mass significantly reducing the need for large loop
corrections to achieve its value 125~GeV. This extra tree level
contribution is proportional to $\lambda v \sin{2 \beta}$,
favoring lower value of $\tan\beta$ and larger value of $\lambda$,
where $v=\sqrt{v_u^2 + v_d^2}$, with $v_u$ and $v_d$ are the vevs for 
up-type and down type Higgs doublets, and $\tan\beta$ is the ratio of these
two vevs~\cite{Ellis:1988er,Drees:1988fc,Franke:1995tc,Ellwanger:1993xa,King:1995vk,Drechsel:2016jdg}.
The physical states, in particular, the neutral Higgses are composed of 
both doublet($\rm H_u,H_d$) and singlet component(S) resulting in a wider  
variation of their masses and coupling strength with fermions 
and gauge bosons.
While offering one of the CP even Higgs as the SM-like 125~GeV Higgs, 
the other Higgs boson states can be very light. For instance, 
depending on the parameter space, if $\rm H_1 \sim \hsm$ ($\rm H_2 \sim \hsm$), 
then the $\rm A_1$ (or both $\rm H_1$ and $\rm A_1$) can be very light, even  
much below than 100~GeV~\cite{Djouadi:2008uw,Heinemeyer:2011aa,King:2012is,Christensen:2013dra,King:2012tr,Guchait:2015owa,Guchait:2016pes,Kumar:2016vhm,Cao:2013gba,Vasquez:2012hn,Domingo:2015eea}. 
Due to the presence of various decay channels of non-SM like Higgs bosons, 
such as, 
$b \bar b,~ \tau^+\tau^-, ~WW, ~ZZ, ~gg, ~c\bar c, ~\gamma\gamma$,
and the wide variation of the corresponding BRs, 
the phenomenology of the Higgs sector in the NMSSM is 
very diverse~\cite{Guchait:2015owa}.
Moreover, in particular,
the Higgs to Higgs decays e.g. $\rm H_1 \to A_1 A_1$, along with 
various other decay modes of $\rm A_1$ 
provide more options to probe the NMSSM Higgs sector 
in colliders~\cite{Almarashi:2010jm,Belyaev:2010ka,Almarashi:2011bf,Bomark:2015fga}.

The parameter space sensitivity to the signal, ~\eqref{eq:signal} 
can be realized by looking into the structure of relevant couplings 
involving the
decay channels, ~\eqref{eq:hsmdk} and ~\eqref{eq:a12gg}.    
Assuming $\rm A_1$ is almost singlet like, the simpler form of 
tree level, $\rm H_1$-$\rm A_1$-$\rm A_1$ coupling is given by,
\br
\rm g_{H_1 - A_1 - A_1}:  
\sqrt{2}v\lambda\left\{ \left[\lambda\left(S_{11}\cos\beta+S_{12} \sin\beta \right)\right]+\left[\kappa\left(S_{12}\cos\beta +S_{11} \sin\beta \right)\right]\right\} 
\label{eq:h1a1a1}
\er
where $S_{11}$ and $S_{12}$ are components of $H_d$ and $H_u$ in the 
physical Higgs boson state.
The more general structure of 
the couplings are shown in the Appendix. 
The above expression clearly shows the explicit dependence of 
BR($\rm H_1 \to A_1 A_1$) on $\lambda$ and $\kappa$.
If this decay is kinematically allowed, then with the 
increase of $\lambda$, the corresponding BR goes
up implying more restrictions on $\lambda$ due to the upper bound on the
SM Higgs BR for undetected decay channels, ~\eqref{eq:betalimit}.
Note that in the context of our study $H_1$ is SM-like Higgs boson i.e
$H_1 \sim H$. 

On the other hand, the reason of larger  
decay rate of $\rm A_1$ to di-photon channel
can be attributed to the singlet nature of $\rm A_1$, which leads to
a very suppressed rate of the tree level decay modes, 
$\rm A_1 \to b\bar b, \tau \bar{\tau}$, but an enhancement of the 
di-photon channel via chargino loop.
Presence of a finite amount of Higgsino composition in 
the chargino state which is present in the loop leads to a 
coupling with singlet like $\rm A_1$ causing  
this enhancement~\cite{Guchait:2016pes}(see the Appendix for details).
Hence, the suppression of tree level decay modes to fermions and little 
enhancement of loop induced couplings of $\rm A_1$ with photons results in a 
larger BR for the di-photon decay channel. A detailed study is carried
out to identify the corresponding parameter space which offer this di-photon BR
large~\cite{Guchait:2016pes}. It is observed that a reasonable range
of $\lambda$($\sim$ 0.1 - 0.4) and $\kappa$($\sim$ 0.1- 0.65)
corresponding to a wide range of values of $A_\lambda$ defined to be
the trilinear Higgs soft breaking parameter and 
(lighter chargino masses ($\sim$ 200 - 700~GeV $\sim  \mu_{\text{eff}}$))
can provide large BR($\rm A_1 \to \gamma\gamma$)\cite{Guchait:2016pes}.
While it is non trivial to find the signal of $A_1$, but it appears as an 
unambiguous indication of the NMSSM like model.
Recall, we mentioned that the constraints on 
$\beta$, ~\eqref{eq:betalimit}, 
from ATLAS data, possibly can constrain the model parameters, 
in particular $\lambda$ and $\kappa$.
\begin{figure}[t]
        \centering
        \includegraphics[width=0.45\linewidth]{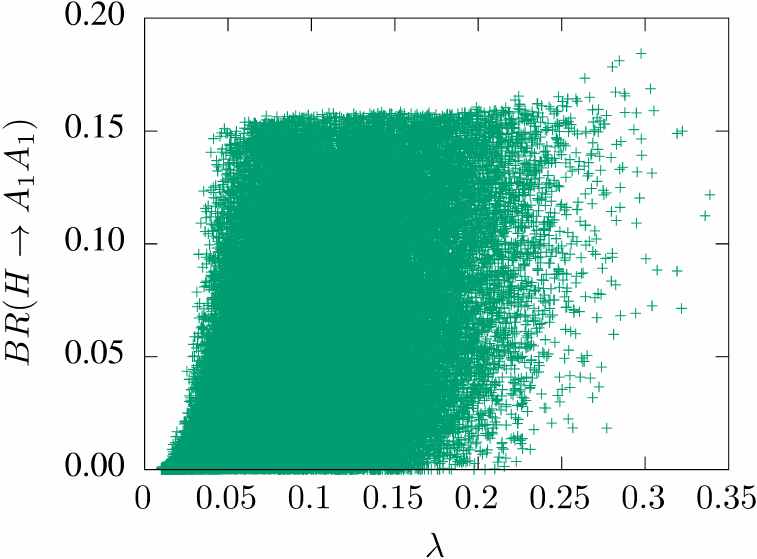}
        \includegraphics[width=0.45\linewidth]{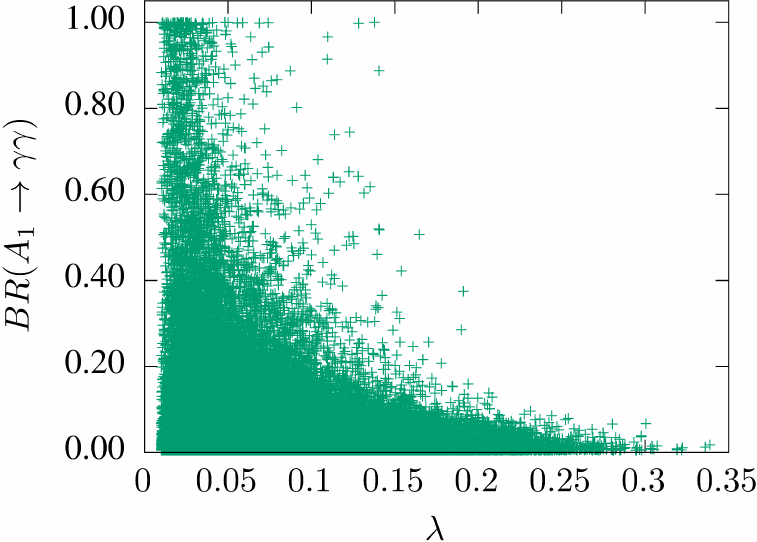}
        \caption{\small{ Allowed regions by all constraints. 
For details see the text.}}
        \label{fig:BRH}
\end{figure}
In order to identify the constrained region of the NMSSM model parameters,   
including constraint given by ~\eqref{eq:betalimit}, we scan over all related 
parameters for a wide range 
using {\tt NMSSMTools}\cite{Ellwanger:2004xm}. The details of the parameters
and the corresponding ranges for scanning are as shown in Table.~\ref{eq:parameter}.
\begin{table}[h!]
\begin{center}
\caption{\small Range of parameters and masses used for 
numerical scan. Energy units are in GeV.}
\begin{minipage}{0.3\linewidth}
        \begin{eqnarray*}
                \lambda          & \in & \left[ 0.01,0.8 \right]\\
                \kappa           & \in & \left[ 0.1,0.8  \right]\\
                A_{\lambda}      & \in & \left[ 100,3000 \right]\\
                A_{\kappa}       & \in & \left[ -20,50   \right]\\
                \mu_{\text{eff}} & \in & \left[ 100,1000 \right]
        \end{eqnarray*}
\end{minipage}
\begin{minipage}{0.3\linewidth}
	\begin{eqnarray*}
		\tan \left( \beta \right) & \in & \left[ 1.5 , 40 \right]\\
		M_{1}     & \in & \left[0,1000 \right]\\
		M_{2}     & \in & \left[100, 1000\right]\\
		M_{3}     & =   & 1500\\
		A_{U_{3}} & \in & \left[-3000,3000\right]
	\end{eqnarray*}
\end{minipage}
\begin{minipage}{0.3\linewidth}
	\begin{eqnarray*}
		A_{E_{3}} & =   & 1500\\
		M_{L_{3}} & =   & 300\\
		M_{E_{3}} & =   & 300\\
		M_{Q_{3}} & \in & \left[800,3000\right]\\
	\end{eqnarray*}
\end{minipage}
\label{eq:parameter}
\end{center}
\end{table}
Here we have used the relations $A_{D_{3}}$= $A_{U_{3}}$, 
 $M_{U_3}$ = $M_{D_3}$ = $M_{Q_3}$, 
 $A_{E_1}$ = $A_{E_2}$ = $A_{E_3}$ and
$M_{L_1}$ = $M_{L_2}$ = $M_{L_3}$.
While scanning parameters, various experimental and theoretical 
constraints incorporated in NMSSMTools are tested.
These constraints include those from flavor Physics, limits on the masses of 
sparticles from LEP, Tevatron and LHC, precision measurements of Higgs 
properties and also the magnetic moment of muon, $g_\mu-2$. 
It is to be noted that the constraint from the measurement of dark 
matter relic density is not taken into consideration.
Fig.~\ref{fig:BRH} presents the allowed range of BR($\rm H\to A_1A_1$)(left)
and BR($\rm A_1 \to \gamma\gamma$)(right) for various values 
of $\lambda$ including 
ATLAS constraint, \eqref{eq:betalimit}. Notice that the higher values of
$(\lambda \gsim$ 0.35), which enhance the $\rm H_1$-$\rm A_1$ - $\rm A_1$
coupling(\eqref{eq:h1a1a1}), and hence the corresponding BR, 
are not favored primarily due to Higgs data, ~\eqref{eq:bsmbr}. 
Remarkably, for the same range of $\lambda$, the
{\BRAG} is found to be large, even as large as $\sim$ 100\%.
We observed that the impact of ATLAS constraint, ~\eqref{eq:betalimit}
is not significant.
\begin{figure}[t]
	\includegraphics[width=0.5\linewidth]{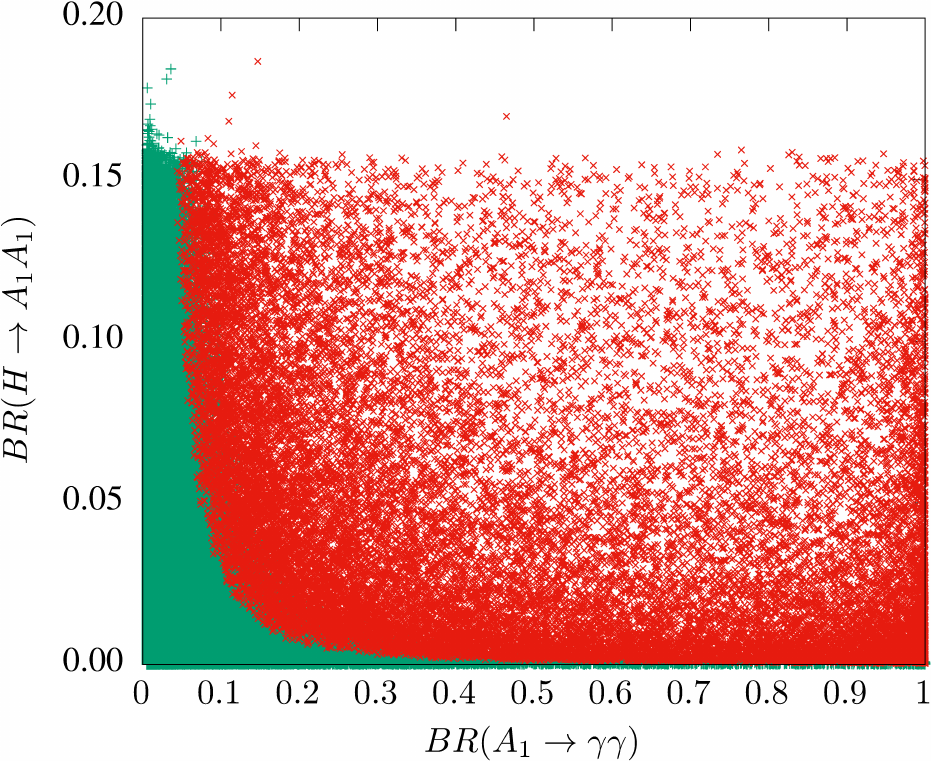}
	\centering
\caption{\small Allowed range of $ \text{BR} \left(\rm H_1 \rightarrow \rm A_1 A_1 \right)$ 
and $ \text{BR} \left(\rm A_1 \rightarrow \gamma \gamma \right)$. 
Red(Green) region is excluded(allowed) by the limit 
on $\beta$, ~\eqref{eq:betalimit}. }
	\label{fig:BRcorr}
\end{figure}
However, we observe an interesting correlation of these 
two BRs, which is
demonstrated in Fig.~\eqref{fig:BRcorr}.
The entire shaded region is excluded by all constraints including the
ATLAS limit, ~\eqref{eq:betalimit}, except the 
the green region which remains unconstrained. 
The notable feature of this plot is that the higher values 
of BR($\rm A_1\to \gamma\gamma$), even though predicted
by the model are not favored by the current ATLAS data.
For a larger range of BR($\rm H_1 \to A_1 A_1$),
smaller values of BR($\rm A_1 \to \gamma\gamma$) are allowed and vice-versa.
It is to be emphasized here that the restrictions of these two BRs along
with their correlations are expected to affect future searches of 
light $\rm A_1$ following these two decay channels.   

\section{Signal and Background}
\label{sec:sandb}
As discussed before, we focus on the signal final state consisting of
at least three photons and one or more leptons along with missing 
transverse energy $\MET$, see \eqref{eq:signal}.
arising via the associated production of SM Higgs 
following the decays, ~\eqref{eq:hsmdk} and \eqref{eq:a1gg}. The hard lepton 
comes from the W decay or semileptonic decay of top quark.
In this context, we briefly discuss about the SM Higgs production cross
section corresponding to our channels, ~\eqref{eq:signal}.
Currently, the $t\bar t H$ cross section in proton-proton collision
is well understood upto next to leading(NLO) order including  
both QCD and EW correction.
The QCD processes contribute dominantly to NLO correction~\cite{Dawson:2003zu,Dawson:2002tg,Reina:2001sf}, where as the 
size of the NLO electro-weak(EW) correction is about $\sim$ 1.7\%
~\cite{Frixione:2015zaa,Frixione:2014qaa,Yu:2014cka}.
The total $t\bar t H$ cross section up to QCD NLO plus EW is found to be about 
500 fb with K-factor 1.25~\cite{deFlorian:2016spz}.
The dominant uncertainty is about $\sim$10\% arising mainly due to 
the variation of QCD scales. A fixed-order computation of the NNLO 
QCD correction is not easy to perform due to the technical difficulties. 
However, attempts 
are made to compute the NNLO QCD correction isolating a particular 
class of higher order diagrams, for  instance, taking into account 
the effect of soft gluon emission beyond 
NLO~\cite{Broggio:2015lya,Kulesza:2015vda}. It is found that this 
approximate NNLO correction reduces the systematic uncertainty in 
the total $t\bar t H$ production by a sizable amount($\sim$5-6\%). 
However, in our estimation we use only NLO QCD+EW corrected 
total  $t \bar{t} H$  cross section. 
We take into account the production of SM Higgs in association with 
a single top, $ t H + \bar{t} H$. 
At LO, $t H$ production takes place via both 
t-channel and s-channel Feynman diagrams in 4 flavor scheme (4FS) as well as
5FS\cite{Demartin:2015uha}.
In 4FS, the final state contains a b-quark or partons in association 
with $t H$ pair, where as in 5FS, the gluon-bottom quark fusion
produces the final state $t W^{\pm} H$ and $tHq$ at the leading order.
Up to NLO level, the t-channel, s-channel and $W^{\pm}$ associated production 
can be distinguished in 5FS case, but in 4FS they interfere at higher 
order level, however, the effect of interference is too small to be 
significant.
Since, we require one lepton in the final state, in our simulation we 
estimate contribution from all channels.
In the 4FS the NLO $tH + \bar t H$ cross section is estimated to be 
67.4 fb with an uncertainty about 1\% due to the top mass and bottom 
quark mass uncertainty\cite{deFlorian:2016spz}, where as in 5FS, 
the NLO cross section is about 74.4 fb with a K-factor 
1.2\cite{deFlorian:2016spz,Demartin:2015uha}.
In our simulation we consider contribution due to the 
$tHq$ productions in 5FS, and the s-channel production corresponding 
to the final
states $tHb$ and $tHW^\pm$.   
The theoretical prediction of the $W^{\pm} H,ZH$ production is currently 
available up to NNLO QCD including EW correction with an uncertainty at 
the level of few percent.
The precise estimations of the total 
$W^{\pm} H$ and ZH production at 13~TeV are found to be  
1.37 pb and 884 fb respectively
at NNLO including both QCD and EW corrections~\cite{deFlorian:2016spz}.

The dominant SM background contribution are due to the processes:
\br
pp \to W\gamma\gamma,~ Z\gamma\gamma,~ t\bar t \gamma\gamma, ~
t\bar t \gamma\gamma\gamma,
\label{eq:bgpro}
\er
where lepton originates from  W or semi-leptonic decay of top quark. 
In all the above hard processes, more number of hard  
$\gamma$ may appear radiatively from the initial states. 
In addition lepton can fake also as a photon.  
We also checked the background contribution from 
$t \bar t \hsm(\to \gamma\gamma)$ and found it to be very small to consider 
due to very tiny BR($\hsm \to\gamma\gamma$) decay.
\begin{table}
	\begin{center}
\caption{\small Production cross sections for signal and background processes}
       	\label{tab:sigma}
		\begin{tabular}{|c|c|c|}
			\hline
Process & $\sigma$(fb) & Source\\
			\hline
$t \bar{t} H$ & 507  &  \cite{deFlorian:2016spz}\\
$W^{\pm} H$ & $1420$ & \cite{deFlorian:2016spz}\\
$Z H$ & 884 & \cite{deFlorian:2016spz} \\
$t H q$ (5FS) & 72.71 & {\tt MG5NLO} \\
$t H b$ (s-channel) & 2.82 & {\tt MG5NLO} \\
$t H W^\pm$ (s-channel) & 15.17 & \cite{deFlorian:2016spz}\\
			\hline
$W^{\pm} \gamma \gamma$ & 407 & {\tt MG5NLO} \\
$Z \gamma \gamma$ & 257 & {\tt MG5NLO} \\
$t \bar{t} \gamma \gamma \gamma$ & 0.67 & {\tt MG5LO} \\
$t \bar{t} \gamma \gamma$ & 13.64  & {\tt MG5NLO} \\
			\hline
		\end{tabular}
	\end{center}

\end{table}
The production cross sections for some of the processes described above 
are computed using {\madgraph}({\tt MG5NLO}) \cite{Alwall:2014hca}
and presented in Table~\eqref{tab:sigma}, which are subject to the following
kinematic cuts on the transverse momentum and rapidity
of leptons, photons and jets:
\br
p_T^{\ell,\gamma} > 10~GeV,~ |\eta_{\ell,\gamma}|<2.5 \\ \nn  
p_T^j \ge 20~GeV,~ |{\eta}_j| <5.
\label{eq:cscuts}
\er
The NNPDF23LO parton distribution function \cite{Ball:2010de} are chosen
to provide input parton flux and setting factorization and 
renormalization scales to $\sqrt{\hat{s}}$, where $\hat s$ is the
energy in the parton center of mass frame. The cross section of 
$t\bar t\gamma\gamma$ is obtained at the NLO multiplying the k-factor
1.35 \cite{Alwall:2014hca} with LO cross section given by MadGraph.
The cross sections for the processes $t\bar t H$, $W^\pm H$ and $ZH$ are 
taken from ref.~\cite{deFlorian:2016spz}.The simulation is performed generating the 
matrix element using {\madgraph} \cite{Alwall:2014hca}, and then for showering and hadronization 
passed through {\pythia} \cite{Sjostrand:2006za}. Events are then stored in  
HepMC format using 
{\hepmc}{\cite{Dobbs:2001ck}} in order to use  
{\delphes}{\cite{deFavereau:2013fsa}} to take into account of detector effects.
In Delphes simulation, we provide inputs through CMS data card 
setting, but changing the photon isolation criteria in order to implement 
our strategy for its selection, as described later. 
We have also verified our results using ATLAS card and found its effect 
not to be very different. In the following, we describe briefly about the 
selection of objects 
in the simulation.

\begin{itemize}	
	
	\item Lepton Selection: Leptons are reconstructed parameterizing the reconstruction 
	efficiencies as a function of both energy and momentum, and the final momentum
	obtained by a Gaussian smearing of the initial momentum. Both electrons and 
	muons are selected, subject to cuts on the transverse momenta 
	($p_T^\ell$) and pseudo rapidity ($\eta^l$) as,
	\br
	p_{T}^{\ell} \ge 20 { ~GeV}, ~|\eta^\ell| \le 2.5, \ \  (l = e, \mu),
	\label{eq:lcut}
	\er
	Restriction on $\eta^\ell$ is due to the tracker coverage in the detector. The 
	cleanliness of the lepton is ensured  
	by measuring the hadronic activities around the lepton 
	direction, requiring the total transverse energy,
	\br
	E_{T}^{AC}(\ell)< 0.12~p_T^{\ell},
	\er
	where $E_T^{AC}(\ell)$ is 
	the scalar sum of transverse energies of all 
	particles with minimum transverse momentum 0.5 GeV around the lepton 
	direction within a cone size of $\Delta R=0.5$.

	\item
	Photon selection: In the Delphes, the genuine photons and electrons, 
	which reach the electromagnetic calorimeter
	without any track - faking as photon,
	are considered. The conversions of photons into electrons and
	positrons pairs are neglected. The photons are selected with cuts,
	\br 
	p_T^{\gamma} > 20,15,\ \  |\eta_{\gamma}| <2.4, 
	\label{eq:gammacut}
	\er
	where, cut on leading and sub-leading photons are 20~GeV and 
	15~GeV respectively.
	Due to the presence of multiple photons in the signal events, 
	isolation of it is checked in such a way that not many genuine 
	photons are missed.
	In this regard, we closely follow the 
	strategy adopted by ATLAS~\cite{Aad:2015bua}
	to select isolated photons.
	First, we estimate the total transverse momentum($E_T^{AC}$) of all
	particles within a region, $\Delta R < 0.4$, around the photon 
	direction. In the next step, the $E_T^{AC}$ is corrected 
	by subtracted out the $p_T$ of a genuine photon, if it  
	is found within an annulus region 0.15$<\Delta R<$0.4 around the 
	photon direction. Finally, we require, $E_T^{AC}<$4~GeV 
	for photon isolation.       
	
	\item 
	The missing transverse energy is estimated from the transverse component 
	of the total energy deposited in the various components of the detector as
	\br
	\vec E{\!\!\!/}_T = - \sum {\vec p_T(i)},
	\label{eq:metcut}
	\er
	$i$ runs over all measured collection in the detector. A cut,
	\br
	\PMET>30~GeV.
	\er
	is applied to select events.
\end{itemize}

\begin{table}
	\begin{center}
		\caption{\small Cross sections($\sigma$) after each set 
of cuts normalizing it by acceptance efficiencies for three values
of $m_{A_1}$(in GeV).}
		\label{tab:sigevt}
		\begin{tabular}{|c|c|c|c|}
			\hline
& $t \bar t H$           & $W^{\pm}H$     & $ZH$ \\
$\sigma(fb) \rightarrow$     &     507.1   &       1420    &      883.9       \\
\hline
& $m_{A_1}$ & $m_{A_1}$ & $m_{A_1}$ \\
Selection & 10~~~~~~~~30~~~~~~~~60  &\ \ \  10~~~~~30~~~~~60& 10~~~~~30~~~~~60\\ \hline
$N_\gamma \ge 3$       &{$105.8$}\ \ \ \  {$142.7$}~~~{$192.8$} & {$355.0$}\ \ \ {$429.7$} \ \ {$620.6$} & {$17.6$}~~{$20.8$}~~{$29.2$} \\
$N_\ell \ge 1$         &{$26.0$} \ \ \ \  {$34.3$}~~~{$47.6$}   & {$52.6$} \ \ \ {$65.6$}  \ \ {$92.5$} & {$14.4$}~~{$17.2$}~~{$24.0$} \\
$\PMET \ge 30$         &{$22.2$} \ \ \ \  {$29.6$}~~~{$40.8$}   & {$36.3$} \ \ \ {$47.0$}   \ \ {$65.0$} & {$0.9$}~~{$1.2$}~~{$1.8$} \\
$m_{\ge 3 \gamma}<130$ &{$21.5$} \ \ \ \  {$28.9$}~~~{$40.3$}  & {$35.9$}   \ \ \ {$46.6$}   \ \ {$64.8$} & {$0.9$}~~{$1.2$}~~{$1.7$} \\ \hline
\end{tabular}
\end{center}
\end{table}

\begin{table}
	\begin{center}
\caption{\small Same as Table ~\eqref{tab:sigevt}, but for $tH$ process.}
\label{tab:thevt}
		\begin{tabular}{|c|c|c|c|}
			\hline
			                    & $t H b$                                                & $t H j$           & $t H W^{\pm}$     \\ 
			$\sigma(fb) \rightarrow$           & {$2.82$}                   & {$72.71$}                    &       {$15.17$}       \\ \hline
& $m_{A_1}$ & $m_{A_1}$ & $m_{A_1}$ \\
			Selection                  & 10 ~~~~ 30 ~~~~ 60                                         & 10 ~~~~ 30~~~~ 60    & 10~~~~ 30~~~~ 60    \\ \hline
			$N_\gamma \ge 3$       & {$0.74$}~~{$1.04$}~~{$1.34$} & {$19.39$}~~{$25.77$}~~{$34.25$} & {$3.38$}~~{$5.34$}~~{$6.74$} \\
			$N_\ell \ge 1$         & {$0.09$}~~{$0.13$}~~{$0.18$} & {$2.48$}~~~{$3.51$}~~~{$4.60$}  & {$0.90$}~~{$1.42$}~~{$1.76$} \\
			$\PMET \ge 30$         & {$0.07$}~~{$0.09$}~~{$0.13$} & {$1.91$}~~~{$2.68$}~~~{$3.51$}  & {$0.75$}~~{$1.21$}~~{$1.45$} \\
			$m_{\ge 3 \gamma}<130$ & {$0.07$}~~{$0.09$}~~{$0.13$} & {$1.90$}~~~{$2.66$}~~~{$3.47$}  & {$0.73$}~~{$1.19$}~~{$1.43$} \\ \hline
		\end{tabular}
	\end{center}
\end{table}

\begin{table}
	\begin{center}
		\caption{\small Background cross section after each set 
of cuts normalizing it by acceptance efficiencies.}
		\label{tab:bgevt}
		\begin{tabular}{|c|c|c|c|c|}
\hline
& W$\gamma\gamma$  & $Z\gamma\gamma$ & $t\bar t \gamma\gamma$ & $t\bar t\gamma\gamma\gamma$ \\ 
$\sigma(fb) \rightarrow$ &    407.0    &    257.0  &    13.64  &    0.670 \\ 
\hline
Selections & & & & \\			
			$N_\gamma \ge 3 $      & {$1.11$}  &  {$0.16$}  & {$0.12$}  & {$0.11$} 			    \\
			$N_{\ell}\ge 1$        & {$0.089$} &  {$0.081$} & {$0.024$} & {$0.028$} \\
			$\PMET > 30$           & {$0.058$} &  {$0.002$} & {$0.021$} & {$0.024$} \\
			$m_{\ge 3 \gamma}<130$ & {$0.026$} &  {$0.001$} & {$0.007$} & {$0.007$} \\  \hline  
		\end{tabular}
	\end{center}
\end{table}

Simulating signal and background processes 
implementing all selection cuts as described above, we estimate 
the signal significance for 
few integrated luminosity options, ${\cal L}=$ 100,~300,~1000 $\rm \invfb$.
Before presenting the signal sensitivity, we discuss the effect of 
selection cuts showing event summary for both the signal
in Tables.~\eqref{tab:sigevt}, ~\eqref{tab:thevt} and backgrounds in 
~\eqref{tab:bgevt}. 
These tables show the cross sections times efficiency
after each set of selection cuts as shown in the 1st column.
The total production cross
sections ($\sigma(\text{fb})$) are shown 
for each of the processes. Three benchmark choices 
of $m_{A_1}=10$~GeV, 30~GeV and 60~GeV 
are considered for the sake of presenting signal rates.  
Notice that, 
the photon selection cuts(~\eqref{eq:gammacut}), reduces the 
background contamination significantly, as clearly seen in 
Table~\eqref{tab:bgevt}.  
The missing transverse energy cut, \eqref{eq:metcut}, is very useful
to suppress the $Z\gamma\gamma$ background, without much reduction in
signal cross sections, except for the ZH case. Since, in both cases, 
due to the absence of genuine sources, the $\MET$
is very soft, and hence affected drastically by a 30 GeV cut.
 
Furthermore, the invariant mass of the photon system
originating from the 
SM Higgs decay of mass 125~GeV via a pair of light $A_1$(\eqref{eq:hsmdk}),
is very effective in isolating backgrounds. 
\begin{figure}[t]
\centering
	\includegraphics[height=7.0cm]{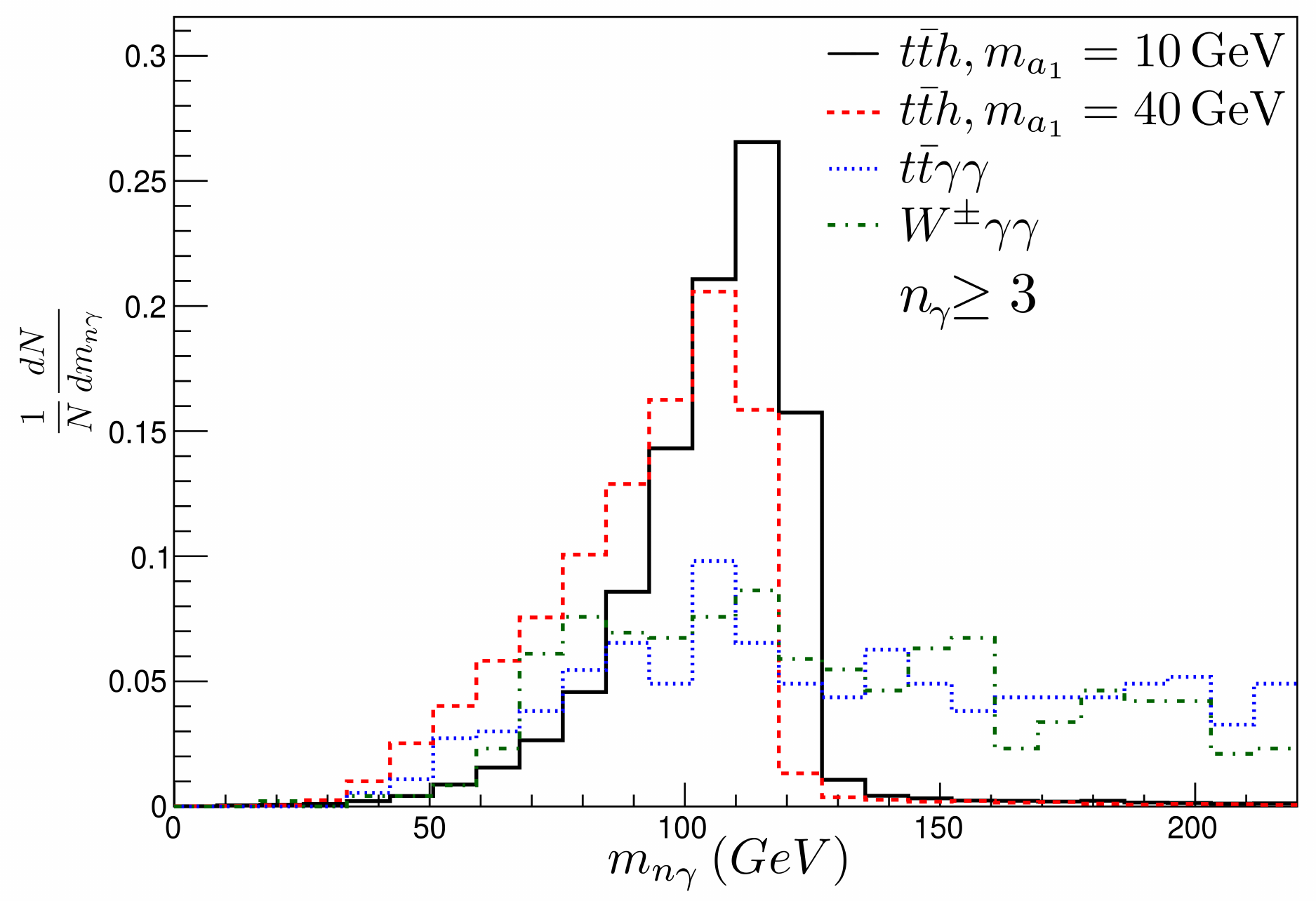}
         \caption{\small Invariant mass of photon system with at 
least three photons subject to photon selection cuts.}
	\label{fig:csmu}
\end{figure}
Fig.\eqref{fig:csmu} demonstrates the invariant mass spectrum 
of the photon system having at least three photons with selection 
cuts, ~\eqref{eq:gammacut}, and 
for the values of $m_{A_1}=$10~GeV and 40~GeV.
For the sake of comparison, in the same figure, the dominant 
backgrounds due to 
$W\gamma\gamma$ and $t\bar t \gamma\gamma$ are also presented.
This invariant mass 
is expected to be bounded by the mass of the Higgs boson, 
and it is clearly observed in Fig.\eqref{fig:csmu}. Obviously, a cut 
on the photon invariant mass $m_{n_{\gamma}\gamma}\lsim 130$~GeV
($n_{n_{\gamma}}\ge$ 3), 
shows an impressive 
discrimination between background and signal events,
without affecting the latter too much, as presented in Tables
~\eqref{tab:sigevt}, ~\eqref{tab:thevt} and ~\eqref{tab:bgevt}. 
Eventually, the total background cross section 
remains to be 0.041~fb, where almost 70\% contribution is due to 
the $W\gamma\gamma$ process. In case of the signal, $\rm W\hsm$ 
is the leading source followed by $tt \hsm$, where as the contributions 
due to $\rm Z\hsm$ and $\rm t\hsm$ are very tiny.
Note that, while presenting the signal cross sections  
in Table ~\eqref{tab:sigevt} and ~\eqref{tab:thevt} the product of 
two BRs, ~\eqref{eq:hsmdk}, 
and ~\eqref{eq:a1gg}, which is essentially $\beta$, is not taken 
into account. 
   
In Table ~\eqref{tab:signi}, we present cross sections for 
the signal sensitivity for three values of $m_{A_1}$ along with the  
background cross section.  
Here the final signal cross sections are obtained by taking into account the 
depletion due to the BRs of Higgs and $\rm A_1$ in terms of  
$\beta$, ~\eqref{eq:beta}, which is restricted to be 
$\beta \lsim 10^{-3}$, ~\eqref{eq:betalimit}. 
We choose limiting value of $\beta = 10^{-3}$
to present signal significance in a model independent way.
Three integrated luminosity options are considered,
${\cal L}=$100,~300,~1000~$\rm\invfb$ to present the discovery 
potential of $\rm A_1$.
Signal sensitivity for lower masses of $\rm A_1$ is 
moderate for 100$\invfb$ integrated luminosity, 
while for higher masses it is quite high. 
For higher luminosity options, 
the entire considered mass range of $\rm A_1$ is easily detectable  
with significance 
more than 5$\sigma$. 
Although results are presented for three choices of
$m_{A_1}$, however, conclusion remains same for the whole range of 10-60 GeV. 
The signal sensitivity degrades by 50\% for a factor of two depletion of 
the limiting value of the combined branching ratio $\beta$.   
Note that this proposed search strategy does not work for 
lower masses ($\lsim$ 10 GeV) of $\rm A_1$, where one requires to develop 
different methodology due to the poor isolation of soft  
photons originating from very light $\rm A_1$ decay~\cite{Dobrescu:2000jt}. 
Finally, as demonstrated, the discovery potential of 
light $\rm A_1$ is quite high, even for 100$\invfb$ integrated 
luminosity option, 
which is expected to be achieved at the end of the next year LHC operation. 
Discovery of light $\rm A_1$, not only establish the new physics signal, 
but can also be an indication of the NMSSM model.
    
\begin{table}
	\begin{center}
		\begin{tabular}{|c|c|c|c|}
			\hline
$m_{A_1}$ 
&  Signal CS(fb) & Bkg CS(fb) & S/$\sqrt{B}$   \\
(GeV)& & &  ${\cal L}$($\invfb$)  \\
&& & 100~~~~~~~~300~~~~~~~~1000\\
\hline
10 & 0.061 & &3.01~~~~~~~~5.21~~~~~~~~~9.51 \\
30 & 0.081 &0.041 &3.97~~~~~~~~6.88~~~~~~~~12.57 \\
60 & 0.111 & &5.51~~~~~~~~9.53~~~~~~~~17.41 \\
			\hline
		\end{tabular} 
		\caption{\small Signal significance for three values of $m_{A_1}$ with $\beta = 10^{-3}$ (see ~\eqref{eq:beta}).}
		\label{tab:signi}
	\end{center}
\end{table}
\section{Summary} \label{sec:conclusion}
The discovery potential of light pseudo scalar 
Higgs boson, $\rm A_1$ of mass less than the half of the SM-like Higgs boson  
is investigated   
at the LHC Run 2 experiment. 
The light Higgs bosons are predicted
by the NMSSM in addition to the SM-like Higgs boson of mass
125~GeV. In this model,
the singlet like light pseudo scalar Higgs boson 
is found to be decaying to a pair of photons 
with a substantial BR.
Numerous studies are already carried out 
by CMS and ATLAS collaborations to search for these light 
Higgs boson in various final states. 
In the absence of any signal, 
limits on the event rates are presented, and those can be translated
to constrain the cross section times BRs corresponding to that final state
for a given model framework.
The recent published results 
by ATLAS in the context of new physics searches with the final state 
consisting of at least three photons impose a stringent constraint
on the combination of BR($\rm H \to A_1 A_1$) and 
BR($\rm A_1 \rightarrow \gamma\gamma$), defined to be $\beta$ in 
~\eqref{eq:beta}. 
In the context of the NMSSM model, 
a significant range of BR($\rm H \to \rm A_1 A_1$) and BR($\rm A_1 \rightarrow \gamma\gamma$)
are excluded for a moderate range of $\lambda$, a very sensitive 
parameter related to these decay modes. 
Moreover, 
we observe that higher range of $\lambda \gsim 0.35$ is disfavored
for the parameter space where the SM-like Higgs is kinematically
allowed to decay to a pair of $\rm A_1$. 
We explore the detection prospect of $\rm A_1$ exploiting its di-photon decay 
channel focusing the final state having at least three 
photons accompanied with one or more hard leptons and missing 
transverse energy. The rate of signal event production 
is limited by $\beta$~\eqref{eq:betalimit}, due to ATLAS data, 
and is taken into 
account in our simulation. Carrying out a detailed simulation for 
signal and background processes including 
detector effects, and considering the limiting value of
$\beta \lsim 10^{-3}$, we present model independent significance for 
three integrated luminosity options. 
The detection prospects of light  
pseudo scalar Higgs boson for the above mass range is modest for 
100~$\rm \invfb$ integrated luminosity. However, 
with high integrated luminosity option, ${\cal L}$=300$\rm \invfb$ or
more, the discovery potential is observed to be quite rich with significance
more than 5$\sigma$.
In summary, we conclude that the discovery of $\rm A_1$ in the diphoton channel
will confirm not only the presence of new physics model, but more importantly, 
can be a characteristic signal of the NMSSM. 
  
\section{Acknowledgement}
The authors are thankful to Ushoshi Maitra and Disha Bhatia who had 
collaborated at the initial phase of this work.

\section{Appendix}
{\bf $\rm H_1$-$\rm A_1$-$\rm A_1$ coupling:} 
The mass matrices for CP-even and CP-odd Higgses in the framework of 
the NMSSM can be diagonalized by orthogonal rotation 
matrices $S_{3 \times 3}$ and $P_{3 \times 3}$ respectively.
The transformation from the weak basis states 
$H_i^s = (H_{dR} , H_{uR}, S_R)$ and 
$H_i^a=(H_{dI},  H_{uI}, S_I)$ to mass basis are given in terms
of mixing matrix elements~\cite{Ellwanger:2009dp}
\begin{equation}
H_i = S_{ij} H_i^s,  ~~~~~~ A_i = P_{ij} H_i^a; \ \ \ \ i,j =1,2,3.
\end{equation}
The Higgs couplings are very sensitive to these elements of the mixing matrices 
$S_{ij}$ and $P_{ij}$.

Assuming the $H_1$ is SM-like, i.e singlet component($S_{13}\sim 0$) 
is almost negligible, and $A_1$ state singlet like, 
i.e $P_{12}, P_{22} \sim 0$, the coupling structure of 
$H_1 - A_1 - A_1$ can be written as~\cite{Ellwanger:2009dp}, 
\begin{eqnarray*}
	g_{H_{1} -A_{1} -A_{1}} & \sim & \frac{\lambda^{2}}{2}\left[v_{d}\Pi_{111}^{133}+v_{u}\Pi_{111}^{233}\right]\\
	&  & +\frac{\lambda\kappa}{\sqrt{2}}\left[v_{d}\Pi_{111}^{233}+v_{u}\Pi_{111}^{133}\right]
\end{eqnarray*}
where $ \Pi_{111}^{i33}  =  2S_{i1}P_{13}^{2}$.  For pure singlet state, 
$P_{13} \sim 1$, it turns out, 
\br
g_{H_1-A_1-A_1} \sim
\sqrt{2}v\lambda\left\{ \left[\lambda\left(\cos\beta S_{11}+\sin\beta S_{12}\right)\right]+\left[\kappa\left(\cos\beta S_{12}+\sin\beta S_{11}\right)\right]\right\} 
\er
with $\tan\beta = {v_u}/{v_d}$.

{\bf $\rm A_1$-$\gamma$-$\gamma$ coupling:} This coupling occurs at the loop 
level and for the singlet dominated $A_1$, the chargino contribution is the 
dominant one. 
The partial decay width of $\rm A_1 \to \gamma\gamma$ is 
given as~\cite{Spira:1995rr,Spira:1997dg},
\begin{equation}
\Gamma(A_1 \rightarrow \gamma \gamma)= 
\frac{G_F \alpha_{em}^2 M_{A_1}^3}{32 \sqrt 2 \pi^3}
\left |\sum_f N_c ~e_f^2 ~g_f^{A_1}~ A_f(\tau_f) ~ + ~
\sum_{\tilde \chi^{\pm}_i} ~ g_{\tilde\chi^\pm_i}^{A_1} ~ A_{\tilde \chi^{\pm}_i}
(\tau_{\tilde \chi^{\pm}_i}) \right|^2.
\label{eq:a1gg}
\end{equation}
Here $\rm N_c$ is the QCD color factor, $e_f$ is the electric charge of the fermions $(f)$,
$A_x (\tau_x)$ are the loop functions given by,
\begin{equation}
A_{x} (\tau_x)=\tau_x \left(\sin^{-1} 
\frac{1}{\sqrt \tau_x} \right )^2,   \ \  \tau_x = \frac{4  M^2_{x}}{M_{A_1}^2};\ \  x = f,~\tilde\chi^\pm_i. 
\end{equation}
Here $g_f^{\rm A_1}$ are the couplings of $\rm A_1$ with the heavier 
fermions (f = t, b) whereas  $g_{\tilde \chi^{\pm}}^{\rm A_1}$ represents 
its couplings with the charginos, these are given by 
\cite{Ellwanger:2009dp},
\begin{align}
g_{u}^{A_1} &= -i \frac{m_u}{\sqrt 2 v \sin \beta} P_{12}, \\ \nonumber  
g_{d}^{A_1} &=  i \frac{m_d}{\sqrt 2 v \cos \beta} P_{11}, \\ \nonumber
g_ {{\tilde \chi_i^{\pm}} {\tilde \chi_j^{\mp}} {A_1}} 	&= \frac{i}{\sqrt 2} \left [ { \lambda} 
P_{13} U_{i2} V_{j2} - 
{g_2} (P_{12} U_{i1} V_{j2} +P_{11} U_{i2} V_{j1})  \right]
\label{eq:coup}
\end{align}
Here U and V are the chargino mixing matrices.
In the pure singlet limit of $\rm A_1$ the mixing elements $P_{11}$ and $P_{12}$ $\sim 0$ and hence  
the fermion couplings $g_u^{A_1}$ and $g_d^{A_1}$ are very small $\sim 10^{-5}$.
Hence, corresponding fermionic loop contribution in ~\eqref{eq:a1gg} 
are extremely small. On the other hand,  
the presence of Higgsino composition in the chargino state yields a 
favorable coupling with $\rm A_1$ via the singlet-Higgsino-Higgsino 
interaction\cite{Guchait:2016pes}.

\bibliography{paper.bib}
\bibliographystyle{utphys.bst}

\end{document}